\documentclass[%
reprint,
superscriptaddress,
showpacs,
preprintnumbers,
amsmath,amssymb,
aps,
prl,
]{revtex4-1}
\usepackage[colorinlistoftodos]{todonotes}
\usepackage{graphicx}
\usepackage{color}
\usepackage[caption=false]{subfig}
\usepackage{dcolumn}
\usepackage{bm}
\usepackage{hyperref}
\usepackage[export]{adjustbox}
\usepackage{subfig}
\usepackage[mathlines]{lineno}
\usepackage{booktabs}

\newcommand{\one}[1]{\textcolor{black}{#1}}
\newcommand{\two}[1]{\textcolor{black}{#1}}
\newcommand{\three}[1]{\textcolor{black}{#1}}
\newcommand{\four}[1]{\textcolor{black}{#1}}

\begin{document}

\preprint{APS/123-QED}

\title{\one{Controlled} drop generation via ligament \one{extraction from a static or vibrating liquid bath}}

\author{Johnathan Hoggarth}
\affiliation{Mechanical Engineering, Yale University.}
\author{Daniel M. Harris}%
\affiliation{School of Engineering, Brown University.}
\author{John W.M. Bush}%
\affiliation{Department of Mathematics, Massachusetts Institute of Technology.}
\author{Bauyrzhan K. Primkulov}%
 \email{bauyrzhan.primkulov@yale.edu}
\affiliation{Mechanical Engineering, Yale University.}%

\date{\today}

\begin{abstract}
We introduce a simple method for generating droplets at the surface of a liquid bath by rapidly stretching a liquid ligament with a spring-loaded cylindrical probe. By varying the probe radius $a$ and retraction length $L$, we identify three regimes. Overstretching a thin ligament produces multiple drops, while insufficient stretching of a thick ligament yields none. The optimal regime for single-drop generation lies in between. In the single-drop regime, the drop radius scales as $R \sim a^{2/3} L^{1/3}$, consistent with volume conservation of the stretched ligament. This method enables repeatable generation of single droplets \one{(with $<5$\% variation in $R$)} on both still and vibrating baths and of ordered droplet lattices for pilot-wave hydrodynamics experiments.
\end{abstract}

\maketitle

\textbf{Introduction}. The controlled generation of liquid drops is central to a wide range of practical applications, including inkjet printing~\citep{lohseFundamentalFluidDynamics2022}, spray cooling~\citep{kimSprayCoolingHeat2007}, and microfluidics~\citep{annaFormationDispersionsUsing2003, garsteckiFormationDropletsBubbles2006}. In such settings, system performance often depends sensitively on drop size. A particularly demanding example arises in pilot-wave hydrodynamics, where millimetric silicone oil droplets self-propel on the surface of a vibrating silicone oil bath through a resonant interaction with its guiding or pilot wave~\citep{bushPilotWaveHydrodynamics2015, bushHydrodynamicQuantumAnalogs2021}. In these experiments, droplet size strongly influences the bouncing and walking thresholds, walking speed, and access to distinct dynamical regimes~\citep{protiereSelforganizationCapillaryWave2005, protiereParticleWaveAssociation2006, wind-willassenExoticStatesBouncing2013}. Moreover, the emergent statistics arising in both diffraction~\citep{couderSingleParticleDiffractionInterference2006,pucciWalkingDropletsInteracting2018, pucciSingleparticleDiffractionHydrodynamic2025, primkulovDiffractionWalkingDrops2025} and corral~\citep{harrisWavelikeStatisticsPilotwave2013, saenzStatisticalProjectionEffects2017} experiments are highly sensitive to drop size. Generating a single sub-millimetric drop of controlled size \four{on a vibrating bath} is thus critical to reproducible results.

One can generate drops with a very narrow size distribution with a nozzle-based pendant-drop technique, in which a droplet grows quasi-statically and detaches once gravity overcomes capillary pull~\citep{tateMagnitudeDropLiquid1864}. However, in order to generate the submillimetric silicone oil drops used in pilot-wave hydrodynamics~\citep{protiereSelforganizationCapillaryWave2005,protiereParticleWaveAssociation2006,wind-willassenExoticStatesBouncing2013}, one would need to use impractically small nozzles~\citep{harrisPilotwaveDynamicsWalking2015, primkulovTatesLawGeometric2026a}. In early walking droplet experiments, droplets were generated by manually pulling a liquid ligament from the bath surface using a toothpick or similar probe~\citep{protiereParticleWaveAssociation2006, eddiArchimedeanLatticesBound2009}. While simple and inexpensive, the success of this approach depends strongly on user technique and offers limited control over drop size and reproducibility (with a characteristic variation of $23$\% in $R$, see Supplemental Video 1). This technique is advantageous in that it can be easily extended to the generation of compound droplets. For example, \citet{perrardSelforganizationQuantizedEigenstates2014} generated ferrofluid drops encapsulated in silicone oil by withdrawal from a two-layer bath. Subsequent efforts to mechanize this ligament-pulling process improved repeatability, but also highlighted the sensitivity of drop formation to probe immersion depth and retraction speed~\citep{eddiMarcheursDualiteOndeparticule2011}. Later, piezoelectric drop generators\three{, utilizing a similar mechanism to those found in inkjet printers,} enabled substantially improved size precision ($\pm 1$\% variation in $R$)~\citep{harrisLowcostPrecisePiezoelectric2015, ionkinNoteVersatile3Dprinted2018}. However, such devices require careful deceleration of the detached droplet before it reaches the bath, since a high-impact landing can trigger splashing or immediate coalescence. Soft landings have been achieved either by guiding the droplet along a curved surface~\citep{harrisLowcostPrecisePiezoelectric2015}, or by precisely timing a pressure reduction in the piezoelectric device at pinch-off so that the droplet briefly hovers upon detachment~\citep{gabbardDropReboundLow2025}.

Here, we introduce a simple mechanical drop generator that occupies a practical middle ground between manual ligament pulling and piezoelectric dispensers. Constructed from a modified click pen, the device uses latch-mediated spring actuation~\citep{patekLatchmediatedSpringActuation2023} to rapidly retract a cylindrical probe from the bath and stretch a liquid ligament that pinches off into a single droplet. The method is substantially more reproducible than manual techniques, yet far easier to build and operate than piezoelectric generators, as it requires no electronics
By varying the probe radius and retraction distance, we identify regimes of single-drop formation, fragmentation, and ligament retraction. Within the single-drop regime, fixed settings yield droplets of repeatable size, and simple scaling arguments rationalize the observed trends.

\textbf{Experimental Method}. We construct a simple spring-loaded device by attaching a cylindrical probe to a repurposed click-pen mechanism. The probe is positioned such that its base lies at a height corresponding to the maximum equilibrium meniscus rise, $|z_s|$, along the probe (detailed in the Supplemental Materials~\cite{supplemental}). In practice, this condition is achieved through a short iterative procedure in which the probe is lowered to establish contact with the bath and equilibrium wetting, raised until the meniscus caves inward, and then lowered slightly until the meniscus is vertical at the contact line. This protocol yields a reproducible initial meniscus geometry.

\begin{figure*}[t]
    \centering
    \includegraphics[width=\linewidth]{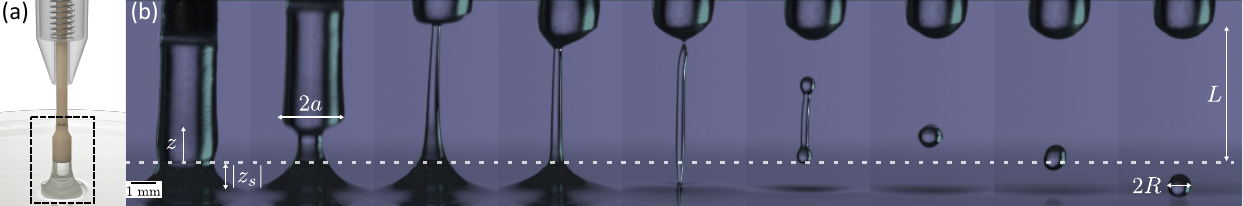}
    \caption{(a)~Schematic of the click-pen droplet generator. (b)~Time sequence of a click-pen retraction that stretches the meniscus beneath a cylindrical probe into a liquid ligament, which then pinches off at both ends to form a single droplet. The first frame corresponds to the onset of retraction ($t=0$). This sequence is shown with equal time intervals of $\Delta t \approx 5.2$\;ms. The sequence corresponds to $L = 7$\;mm and $a = 1.25$\;mm.}
    \label{fig:single_drop}
\end{figure*}

Manually pressing the pen button releases the latch, causing the probe to retract rapidly which stretches the meniscus into a liquid ligament. By adjusting the pen parameters, we independently control the retraction length $L$, probe radius $a$, and characteristic retraction speed (see Fig.~\ref{fig:single_drop}). Additional details on device construction, meniscus calculations, and positioning are provided in the Supplemental Materials~\cite{supplemental}.
 
\textbf{Physical Picture and Phase Diagram}.
We focus on the high-Reynolds-number stretching of the liquid ligament. The spring loading and stiffness are tuned to produce a characteristic retraction speed $U \sim 1$\;m/s. This results in a Reynolds number, $\mathrm{Re}  = UL/\nu \approx 250$, for a kinematic viscosity $\nu = 20$\;cSt and a ligament of length $L=5$\;mm, indicating that the ligament stretching dynamics are inertially dominated.

\begin{figure*}
    \centering
    \includegraphics[width=\linewidth]{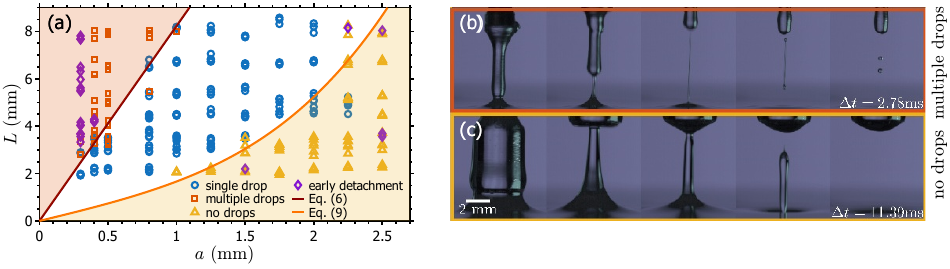}
    \caption{(a)~Phase diagram showing the dependence of drop formation on probe radius $a$ and retraction distance $L$. \one{If a ligament is stretched too far, the Rayleigh–Plateau causes it to pinch off into multiple drops (red regime, bounded by Eq.~\eqref{eqn:RP_boundary}). Conversely, an insufficiently stretched ligament retracts back into the bath without pinch-off (orange regime, bounded by Eq.~\eqref{eqn:g_boundary}). The optimal single-drop regime (white) lies between these two bounds.} Representative time-lapse sequences for the multiple-drop and no-drop regimes are shown in (b) and (c). Purple data points (early detachment), which always lie outside the single-drop regime, denote experiments in which the ligament pinched off from the probe before the probe had traveled the full distance $L$ (see Fig.~S7).}
    \label{fig:phase_diagram}
\end{figure*}

We conducted a sweep of experiments for $L \in [2,9]$\;mm and $a \in [0.3,2.5]$\;mm. Three distinct regimes emerged: formation of a single drop (Fig.~\ref{fig:single_drop}b), ligament break up into multiple drops (Fig.~\ref{fig:phase_diagram}b), and ligament retraction into the bath with no drop formation (Fig.~\ref{fig:phase_diagram}c). We proceed by rationalizing the two boundaries of the single-drop regime, which span two distinct ranges of the Bond number $\mathrm{Bo}=a^2/\ell_c^2$, where $\ell_c = \sqrt{\sigma/\rho g}$ is the capillary length, $g$ is gravitational acceleration, and $\sigma=20.1$\;mN/m and $\rho=950$\;kg/m$^3$ are the surface tension and density of the silicone oil used in our experiments.

\begin{figure}
    \centering
    \includegraphics[width=\linewidth]{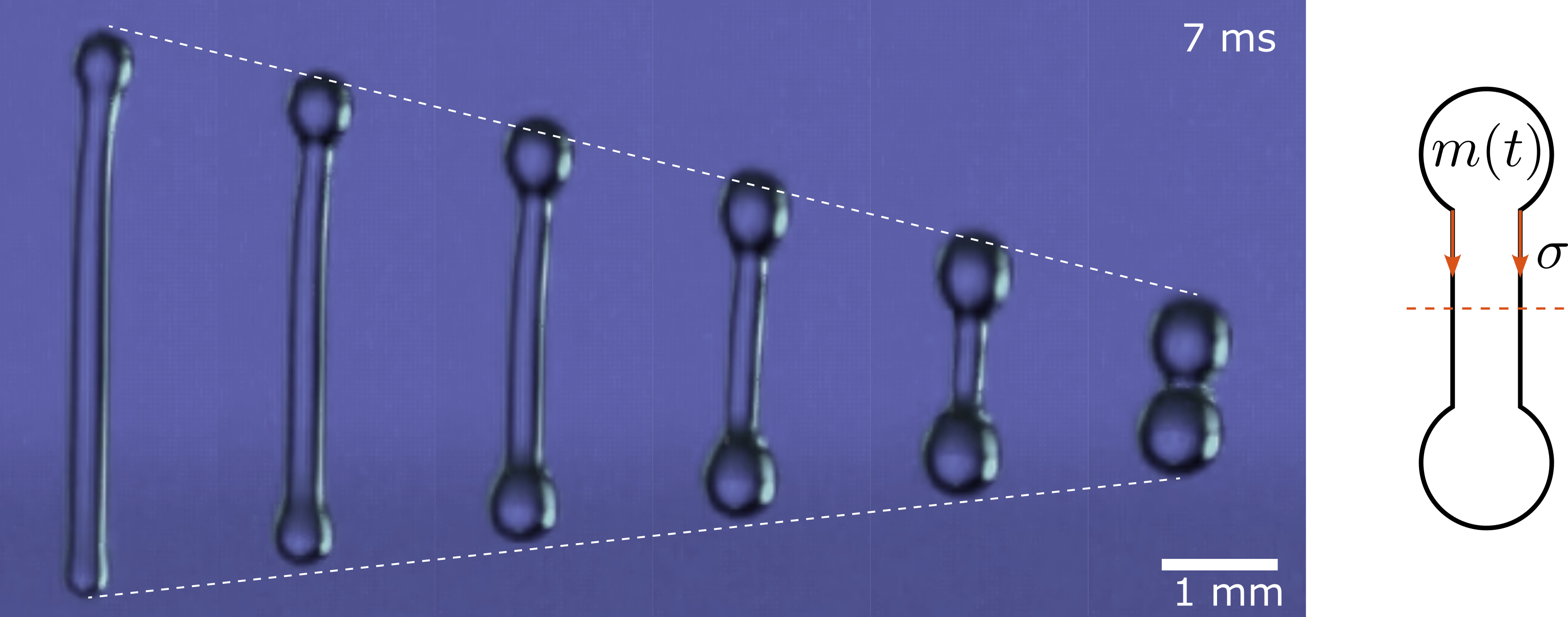}
    \caption{Time sequence images and a schematic of a ligament retracting into a single droplet. As a ligament retracts, its two lobes grow and merge into a single drop. This process takes $\sim7$ milliseconds, and the retraction speed is approximately constant, as indicated by the dashed lines near both ends of the ligament. The slight asymmetry in lobe speeds may be attributed to the entire filament accelerating in response to gravity.}
    \label{fig:ligament_collapse}
\end{figure}

For $\mathrm{Bo} \ll 1$, once the ligament pinches off from both the probe and the bath, surface tension drives its rapid retraction into a single droplet. \one{The stability and collapse of free liquid filaments have been considered in various contexts~\citep{lordrayleighCapillaryPhenomenaJets1879, marmottantFragmentationStretchedLiquid2004, eggersPhysicsLiquidJets2008, notzDynamicsBreakupContracting2004, castrejon-pitaBreakupLiquidFilaments2012, driessenStabilityViscousLong2013, piersonRevisitingTaylorCulickApproximation2020}. Here, we briefly outline the timescale of retraction in order to rationalize the form of our regime diagram (Fig.~\ref{fig:phase_diagram})}. A typical time-lapse of this collapse is shown in Fig.~\ref{fig:ligament_collapse}. The associated timescale can be estimated by considering the force balance on one of the lobes that forms at the ligament ends (see schematic in Fig.~\ref{fig:ligament_collapse}). At the onset of collapse, the ligament has a characteristic length proportional to $L$ and radius proportional to $a$. The momentum balance for the upper lobe may be written as
\begin{equation}
\frac{d}{dt}(m\dot{z}) \approx 2\pi a \sigma + mg,
\label{eqn:ligament_momentum}
\end{equation}
where $m(t)$ is the lobe mass and $g$ is the gravitational acceleration.

As is evident in Fig.~\ref{fig:ligament_collapse}, the ligament retracts nearly at constant speed, so we neglect the inertial term $m\ddot{z}$. Furthermore, we neglect the gravitational term since $2\pi a\sigma/mg = \mathcal{O}(10)$, reducing equation~\eqref{eqn:ligament_momentum} to
\begin{equation}
\dot{m}\dot{z} \approx 2\pi a \sigma.
\label{eqn:red_lig_mom}
\end{equation}
As the lobe retracts, it absorbs liquid from the ligament, so its mass increases according to
\begin{equation}
\dot{m}=\pi a^2 \rho \dot{z}.
\label{eqn:mass_rate_lig}
\end{equation}

Combining equations~\eqref{eqn:red_lig_mom} and \eqref{eqn:mass_rate_lig} gives the characteristic retraction speed
\begin{equation}
u_c=\dot{z}\approx \sqrt{\frac{2\sigma}{\rho a}},
\end{equation}
which is analogous to the classical Taylor–Culick velocity for retracting liquid sheets~\citep{taylorDynamicsThinSheets1959, culickCommentsRupturedSoap1960}.
Since each lobe travels approximately half the ligament length, the corresponding retraction timescale is
\begin{equation}
\tau_c=\frac{L/2}{u_c}\approx \sqrt{\frac{\rho a L^2}{8\sigma}}.
\end{equation}

If this timescale is greater than the inviscid Rayleigh-Plateau timescale~\citep{lordrayleighCapillaryPhenomenaJets1879}
\begin{equation*}
    \tau_\text{RP} \approx 2.91 \sqrt{{\rho a^3}/{\sigma}},
    \label{eqn:RP}
\end{equation*}
then the ligament breaks up into multiple drops before it has the time to retract into a single drop. \one{The choice of $\tau_\text{RP}$ is justified by the low Ohnesorge number of our system, $\text{Oh} = \nu\sqrt{\rho/( \sigma a)} \in [0.09, 0.22] < 1$, which indicates that capillary breakup is primarily resisted by inertial rather than viscous effects.} Comparison between the two time scales yields a linear scaling between $L$ and $a$; specifically, a single drop will form provided 
\begin{equation}
    L/a \lesssim 8.23. \label{eqn:RP_boundary}
\end{equation}
We plot the boundary between the single- and multiple-drop regimes in red in Fig.~\ref{fig:phase_diagram}a. \two{Above this boundary, for sufficiently large aspect ratios $L/a$, the ligament becomes unstable to the Rayleigh–Plateau instability prior to collapse. In this regime, multiple pinch-off events occur along the ligament, producing two or more droplets rather than a single drop (as shown in Fig.~\ref{fig:phase_diagram}b and Supplemental Video 3).} 

\begin{figure}
    \centering
    \includegraphics[width=\linewidth]{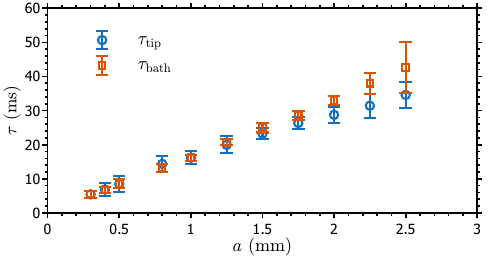}
    \caption{Dependence of the liquid ligament pinch-off times near the probe $\tau_\mathrm{tip}$ and the bath $\tau_\mathrm{bath}$ on the probe radius $a$. The data corresponds to experiments from Fig.~\ref{fig:phase_diagram}a. These times are measured from the start of the probe retraction. Error bars represent the standard deviation across $n=18-30$~trials per probe radius.}
    \label{fig:pinch_time}
\end{figure}

The rationale for the second boundary in Fig.~\ref{fig:phase_diagram}a, which corresponds to $\mathrm{Bo}=\mathcal{O}(1)$, can be understood by considering the difference in the pinch-off dynamics at the tip and at the bath. Our measurements of the pinch-off times near the pen ($\tau_\text{tip}$) and the bath ($\tau_\text{bath}$), measured relative to the onset of retraction, show a weak dependence on the retraction length and speed, but both increase substantially with the probe radius $a$ (Fig.~\ref{fig:pinch_time}). Notably, $\tau_\text{bath}$ increases more rapidly with $a$; consequently, the difference in the duration of the two pinch-off events increases progressively as the probe radius grows. One might anticipate this asymmetry on the basis of the hydrostatic pressure gradient along the filament, which should encourage pinch-off at the top of the ligament while resisting it at its base. For sufficiently large $a$, the ligament therefore pinches off first from the probe, leaving behind a liquid column attached only to the bath.

The subsequent evolution of the column is then governed by a competition between capillary pinch-off at the bath and drainage of the ligament driven by capillarity and gravity. Pinch-off at the bath is a local process, governed by the inertio-capillary timescale
\begin{equation}
    \tau_\mathrm{bath} \sim \sqrt{{\rho a^3}/{\sigma}}.
\end{equation}

The motion of the ligament's lobe is governed by equation~\eqref{eqn:ligament_momentum}, where the gravity term is no longer negligible. As the lobe moves towards the bath, it absorbs the ligament volume $\Omega$, so $\dot{m}\approx\rho\Omega/\tau_\text{drain}$, where $\tau_\text{drain}$ is the drainage time. Furthermore, writing the lobe speed $\dot{z}$ as $L/\tau_\text{drain}$ allows one to approximate equation~\eqref{eqn:ligament_momentum} by $\rho\Omega L/\tau_\text{drain}^2 = 2\pi a \sigma + \rho g \Omega$, yielding the scaling
\begin{equation}
    \tau_\mathrm{drain} \approx \sqrt{\frac{L/g}{2a/(\mathrm{Bo}L)+1}}.
    \label{eqn:tau_drain}
\end{equation}
In the $\mathrm{Bo}\gg1$ limit, equation~\eqref{eqn:tau_drain} reduces to $\tau_\mathrm{drain} \approx \sqrt{{L/g}}$. \one{This gravity-limited regime is similar to those observed in related large Bond number systems, such as fluid motion during cat lapping~\citep{reisHowCatsLap2010}}. In the $\mathrm{Bo}\ll 1$ limit, equation~\eqref{eqn:tau_drain} reduces to $\tau_\mathrm{drain} \approx\sqrt{\frac{\rho a L^2}{8\sigma}}$, which matches the timescale of capillary collapse $\tau_c$, as anticipated.

The ligament fails to pinch off when drainage occurs faster than the capillary breakup at the bath, $\tau_\mathrm{drain} \lesssim \tau_\mathrm{bath}$. Equating these timescales gives the second requirement for single drop formation:
\begin{equation}
    L/a \gtrsim {\mathrm{Bo}}/{2}+ \sqrt{({\mathrm{Bo}}/{2})^2+2}. \label{eqn:g_boundary}
\end{equation}
This scaling rationalizes the upward curvature of the \three{orange} phase boundary at larger probe radii. We plot this prediction directly, with no fitted prefactor, as a boundary of the \three{orange} region in Fig.~\ref{fig:phase_diagram}a. The prediction provides reasonable agreement with the observed boundary.

\begin{figure}
    \centering
    \includegraphics[width=\linewidth]{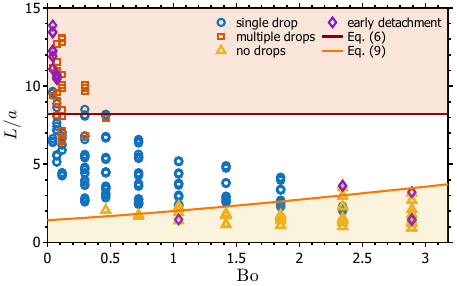}
    \caption{\four{Phase diagram showing the dependence of drop formation on the Bond number $\mathrm{Bo}$ and the aspect ratio of the ligament $L/a$.}}
    \label{fig:phase_Bo}
\end{figure}

Taken together, these two limits define the boundaries of the phase diagram. The single-drop regime lies between them, where both capillary and gravitational effects are relevant. In our experiments, this occurs for probe radii on the order of the capillary length ($a \sim \ell_c$), where the ligament is sufficiently thick that the collapse timescale is shorter than the Rayleigh--Plateau timescale ($\tau_c < \tau_\mathrm{RP}$), allowing a single drop to form, yet slender enough that gravity-driven drainage occurs more slowly than pinch-off at the bath ($\tau_\mathrm{drain} > \tau_\mathrm{bath}$). This combination of conditions defines a controlled window for generating single droplets. \four{Because these competing timescales depend only on ligament geometry and the relative importance of gravity and capillarity, it is illustrative to plot our experimental data using dimensionless parameters. Fig.~\ref{fig:phase_Bo} recasts the data from Fig.~\ref{fig:phase_diagram}a through consideration of a dimensionless framework characterized by the aspect ratio $L/a$ and the Bond number $\mathrm{Bo}$.}

\begin{figure*}
    \centering
    \includegraphics[width=\linewidth]{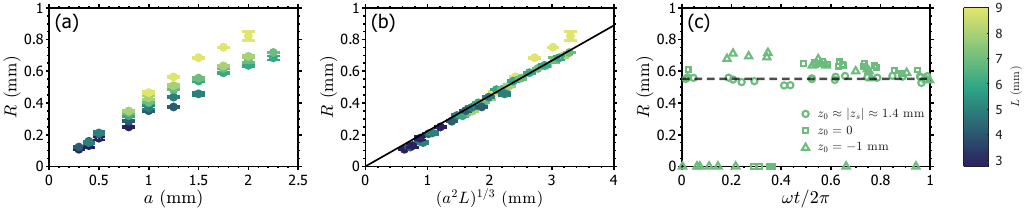}
    \caption{\one{Droplet size scaling and variability for single-drop events. (a)~Drop radius $R$ as a function of probe radius $a$, with marker color representing different retraction lengths $L$. (b)~Drop radius $R$ as a function of $(a^2L)^{1/3}$. As predicted by  Eq.~\eqref{eqn:drop_scaling} the data from (a) collapses onto a linear fit $R = (0.222 \pm 0.005)\left(a^{2}L\right)^{1/3}$~mm. Error bars in (a,b) denote standard deviation across $n=3\text{--}5$ experimental repetitions. (c)~Drop radius $R$ plotted against dimensionless bath vibration phase $\omega t / 2\pi$ ($f=80\,$Hz) for three normalized initial probe positions $z_0/\Delta h$. At $z_0/\Delta h \approx 7.4$, drop size remains consistent with a quiescent bath with $R = 0.55 \pm 0.02$~mm (dashed line); conversely, deeper immersion depths ($z_0/\Delta h = 0, -5.3$) introduce high variability and intermittent failures in drop formation ($R=0$).}}
    \label{fig:drop_size}
\end{figure*}

\textbf{Size of a drop}. When $a$ and $L$ are such that a single drop is generated, its radius $R$ varies with both parameters (see Fig.~\ref{fig:drop_size}a). Specifically, increasing the radius of the cylinder can substantially increase the size of the drop, while pulling the ligament a little further (larger $L$) has a similar but weaker effect. The drop volume $\Omega$ can be deduced from the size of the liquid column at pinch-off. The height of the column should be comparable to $L$, while the radius of the liquid column should be comparable to $a$. Thus $\Omega \sim a^2L$ or
\begin{equation}
    R \sim a^{2/3} L^{1/3}.
    \label{eqn:drop_scaling}
\end{equation}
Rescaling the drop-size data from Fig.~\ref{fig:drop_size}a according to Eq.~\eqref{eqn:drop_scaling} yields a collapse onto a line that corresponds to a prefactor of $0.222 \pm 0.005$ (Fig.~\ref{fig:drop_size}b). 

All of the data reported thus far have been collected on stationary liquid baths. However, in pilot-wave hydrodynamics, a bouncing drop is sustained by vertical vibration of the liquid bath, typically at 50-80 Hz. With vibrational forcing, the vertical position of the probe relative to the bath becomes critical, which we explore in Fig.~\ref{fig:drop_size}c \four{with a fixed probe geometry of $a=1.25~\textrm{mm}$ and $L = 7~\textrm{mm}$}. Experiments were conducted at accelerations just below the Faraday threshold ($\gamma =4.86\,\mathrm{g}$), representing a worst-case scenario in which the vertical bath displacement is maximal ($\Delta h = 0.19$~mm). We plot $R$ versus the dimensionless phase of the bath, $\omega t/2\pi$, where $\omega$ is the bath’s angular frequency. The phase is defined such that $\omega=0$ corresponds to the maximum upward velocity of the bath, with $t=0$ corresponding to the onset of probe retraction. This phase encodes whether the bath is moving upward or downward relative to the probe at the start of retraction.

When the probe begins at the interface ($z_0 = 0$) or is submerged below it ($z_0 = -1~\mathrm{mm}$), droplet formation can intermittently fail, and the drop size becomes more variable (Fig.~\ref{fig:drop_size}c). However, the results obtained for a stationary bath carry over to vibrating baths provided the probe is lifted to a height of $z_0 \approx |z_s|\approx 1.4~\mathrm{mm}$. In this regime, the droplet radius is consistent with that obtained on a quiescent bath, with $R = 0.564 \pm 0.007~\mathrm{mm}$ for the still bath and $R = 0.55 \pm 0.02~\mathrm{mm}$ for the vibrating bath, as shown in Fig.~\ref{fig:drop_size}c, indicating negligible dependence on the bath's phase of vibration. 

\begin{figure}
    \centering
    \includegraphics[width=\linewidth]{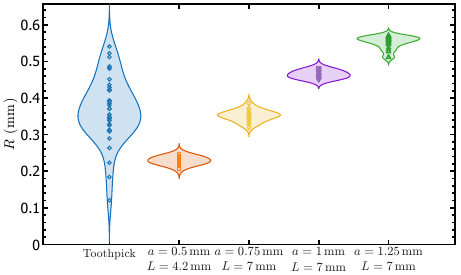}
    \caption{\one{Comparison of distributions of droplet radius R achieved with the manual (toothpick-based) ligament pulling technique and the pen generator with various probe geometries. Shaded regions represent smoothed probability distributions of the drop sizes. The toothpick method yields a mean droplet radius of $R = 0.37 \pm 0.09$~mm, displaying poor reproducibility. In contrast, the pen generator allows the mean droplet radius to be tuned from $R = 0.23$~mm to $0.55$~mm via probe selection, while maintaining an absolute standard deviation of $\le 0.02$~mm and a relative radius variability of $<5\%$ across all probe geometries.}}
    \label{fig:method_comparison} 
\end{figure}

\one{To demonstrate the advantages of the pen-based method over traditional manual techniques, we compared the consistency of droplet formation across a range of probe geometries ($a = 0.5$, $0.75$, $1.0$, and $1.25$~mm) to manual ligament pulling using a standard wooden toothpick (diameter $d \approx 1.6$~mm with a conical taper down to a tip diameter of $\approx 0.5$~mm) (Supplemental Video 1). As shown in Fig.~\ref{fig:method_comparison}, the manual method yields highly variable droplet sizes. In contrast, the pen generator maintains a tight size distribution across all tested geometries. This disparity can be explained by both geometric and mechanical factors. While a conical toothpick is highly sensitive to immersion depth, since any variation alters the meniscus size, the cylindrical geometry of our pen probes, coupled with the ability to control immersion depth, ensures an invariant initial meniscus prior to retraction. Furthermore, the spring-loaded click mechanism guarantees highly consistent retraction speeds and lengths relative to those in manual pulling.}

\begin{figure}
    \centering
    \includegraphics[width=\linewidth]{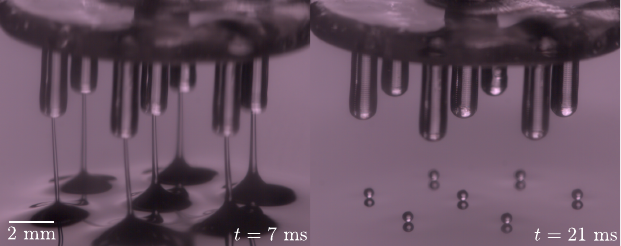}
    \caption{Images of multi-droplet generation using a multi-tip probe. Contact of the probes spaced out by $\lambda_F$ produces a hexagonal lattice of bouncing drops~\citep{eddiArchimedeanLatticesBound2009}.}
    \label{fig:multi_drop}
\end{figure}

Once the parameter space for consistent droplet generation on a vibrating bath was established, we explored whether the device could be extended to “print-in-place” ordered droplet configurations in a single actuation. \one{For a bath with a vibrational frequency of $f = 80$\;Hz, the gravity-capillary dispersion relation $\omega_F^2 = gk_F + (\sigma/\rho)k_F^3$ yields a theoretical wavelength of $\lambda_F = 2\pi/k_F \approx 4.88\text{ mm}$~\citep{kumarLinearTheoryFaraday1996}.} We thus fabricated tips with multiple probes, spaced by approximately one Faraday wavelength. Upon a single actuation of the pen, each tip generated a droplet, resulting in an array of bouncing droplets arranged in a prescribed geometry\two{, as shown in Fig.~\ref{fig:multi_drop} (and Supplemental Video 6)}. \two{Although bouncing droplets are known to form a variety of stable two-dimensional crystal lattices~\citep{eddiArchimedeanLatticesBound2009, simulaDropletTimeCrystals2023} and exhibit rich collective dynamics~\citep{couchmanFreeRingsBouncing2020, thomsonCollectiveVibrationsConfined2020, couchmanStabilityHydrodynamicBravais2022, evansPhaseControlBouncing2026}, assembling these lattices experimentally has remained a laborious manual process.} \two{The demonstration in Fig.~\ref{fig:multi_drop}} highlights the potential of the device to generate ordered droplet arrays through tailored probe geometries.

\textbf{Conclusion}. We have demonstrated a simple mechanical method for generating single droplets by the stretching and spontaneous pinch-off of a liquid ligament from a bath. The resulting phase diagram (Fig.~\ref{fig:phase_diagram}a)~has been rationalized through consideration of the competition between capillarity, inertia, and gravity. We have shown how the method can be made robust and insensitive to bath vibration. Beyond facilitating controlled walker generation, this framework naturally extends to multi-tip geometries and to the generation of multiphase droplets~\citep{perrardSelforganizationQuantizedEigenstates2014}. By transitioning from single- to multi-drop generation, this device offers a ``print-in-place'' capability for the rapid assembly of complex hydrodynamic lattices and the study of many-body pilot-wave interactions.

\textbf{Acknowledgements}.
The authors thank Amir Pahlavan, Saurabh Nath, and Douglas Holmes for insightful discussions. BP and JH gratefully acknowledge support from startup funds provided by Yale University. DMH gratefully acknowledges the financial support of the National Science Foundation (NSF CBET-2123371). JWMB gratefully acknowledges the financial support of the National Science Foundation (CMMI-2154151) and the Office of Naval Research  (No.N00014-24-1-2232).

\bibliography{references.bib}

\end{document}